\documentclass{mn2e}
\usepackage{epsfig}

\def\ngc{{NGC 4051}}

\def\xmm{{\it XMM-Newton}}
\def\chandra{{\it Chandra}}

\def\et{{et al.\ }}

\newcommand{\ls}{\mathrel{\hbox{\rlap{\hbox{\lower4pt\hbox{$\sim$}}}\hbox{$<$}}}}
\newcommand{\gs}{\mathrel{\hbox{\rlap{\hbox{\lower4pt\hbox{$\sim$}}}\hbox{$>$}}}}

\def\H0{{\rm ~km~s^{-1}~Mpc^{-1}}}

\def\et{{et al.}}

\def\deg{^\circ}

\title[X-ray spectrum of \ngc]
{An extended \xmm\ observation of the Seyfert galaxy \ngc. 
II. Soft X-ray emission from a limb-brightened  shell of post-shock gas}
\author[K.A.Pounds \et]
        {K.A.Pounds,
	S.Vaughan 
	\\
Department of Physics and Astronomy, University of Leicester,
Leicester, LE1 7RH, UK\\}

\date{Accepted ; Submitted }
\pagerange{\pageref{firstpage}--\pageref{lastpage}}
\pubyear{2010}
\begin{document}
\maketitle
\label{firstpage}

\begin{abstract}   
An extended \xmm\ observation of the Seyfert 1 galaxy \ngc\ in 2009 revealed a complex absorption spectrum, with a wide range of outflow
velocities and ionisation states. The main velocity and ionisation structure was interpreted in Paper I in terms of a decelerating,  recombining
flow resulting from the shocking of a still higher velocity wind colliding with the ISM or slower moving ejecta. The high sensitivity of the
\xmm\ observation also revealed a number of broad emission lines, all showing evidence of self-absorption near the line cores. The line
profiles are found here to be consistent with emission from a limb-brightened shell of post-shock gas building up ahead of the contact discontinuity. While the
broad emission lines remain quasi-constant as the continuum flux changes by an order of magnitude, recombination continua of several H- and He-like ions are found
to vary in response to the continuum, providing an important key to scaling the ionised flow.    
\end{abstract}


\section{Introduction}

High resolution spectra of the bright Seyfert 1 galaxy \ngc\, obtained by \chandra\  and \xmm\ over the past decade have detected soft X-ray 
absorption lines 
indicating a typical outflow velocity of $\sim$500-600 km s$^{-1}$, with occasional higher blue shifts of up to $\sim$4600 km s$^{-1}$  (Collinge \et\
2001, Ogle \et\ 2004, Pounds \et\ 2004, Steenbrugge \et\ 2009).

A new and much longer \xmm\ observation of \ngc\ in 2009 has now revealed a high resolution soft X-ray spectrum with outflow velocities up to 
$\sim$9000 km s$^{-1}$. In Pounds and Vaughan (2011), hereafter Paper I, we interpreted an observed correlation of outflow velocity and ionisation state in
terms of a decelerating and  recombining flow, resulting from  the shocking of a still higher velocity wind colliding with the ISM or slower moving
ejecta. 

The high sensitivity of the 2009 \xmm\ observation also found several strong and broad emission lines, all showing evidence of self-absorption
near the line cores.
Broad soft X-ray emission lines have previously been reported for \ngc\ (Ogle \et\ 2004, Steenbrugge \et\ 2009), and for several other
Seyfert 1 galaxies (Kaastra \et\ 2002, Costantini \et\ 2007, Smith \et\ 2007), being interpreted in each case as arising in the inner part of the
optical/UV broad line region.  

In the present paper we model the broad emission lines in \ngc\, and consider an alternative origin  in a limb-brightened shell of shocked gas. We note,
furthermore,
that self absorption in the shell also offers a natural explanation for a separate low velocity absorption component, seen across a wide range of ionisation
states and found in Paper I to break the  main velocity-ionisation
correlation. 

A limitation common to many studies of AGN outflows, which bears directly on estimates of mass, energy and momentum, and hence on AGN feedback, arises
from a degeneracy between the radius and particle density of  ionised gas in the flow. Attempts to break this degeneracy by detecting flux-linked
changes in the ionisation state of the absorbing gas have generally been inconclusive. While we find the broad line emission in \ngc\ to remain
quasi-constant, as the continuum flux changes by an order of magnitude, several radiative recombination continua (RRC) of H- and He-like ions do vary,
apparently in response to the continuum. We examine those data to constrain the recombination time of the photoionised gas and thereby scale the
parameters of the flow in \ngc.

\section{Observations and data analysis}

\ngc\ was observed by \xmm\ on 15 orbits between 2009 May 3 and June 15, yielding an overall exposure of $\sim$600 ks for each of the Reflection Grating
Spectrometers, RGS 1 and RGS 2 (den Herder \et\ 2001).    Full details on the timing,  flux levels and X-ray light curves for each satellite orbit of the 
2009 observation are
included in an accompanying paper on the `Rapid X-ray variability of \ngc' (Vaughan \et\ 2011).

\begin{figure}
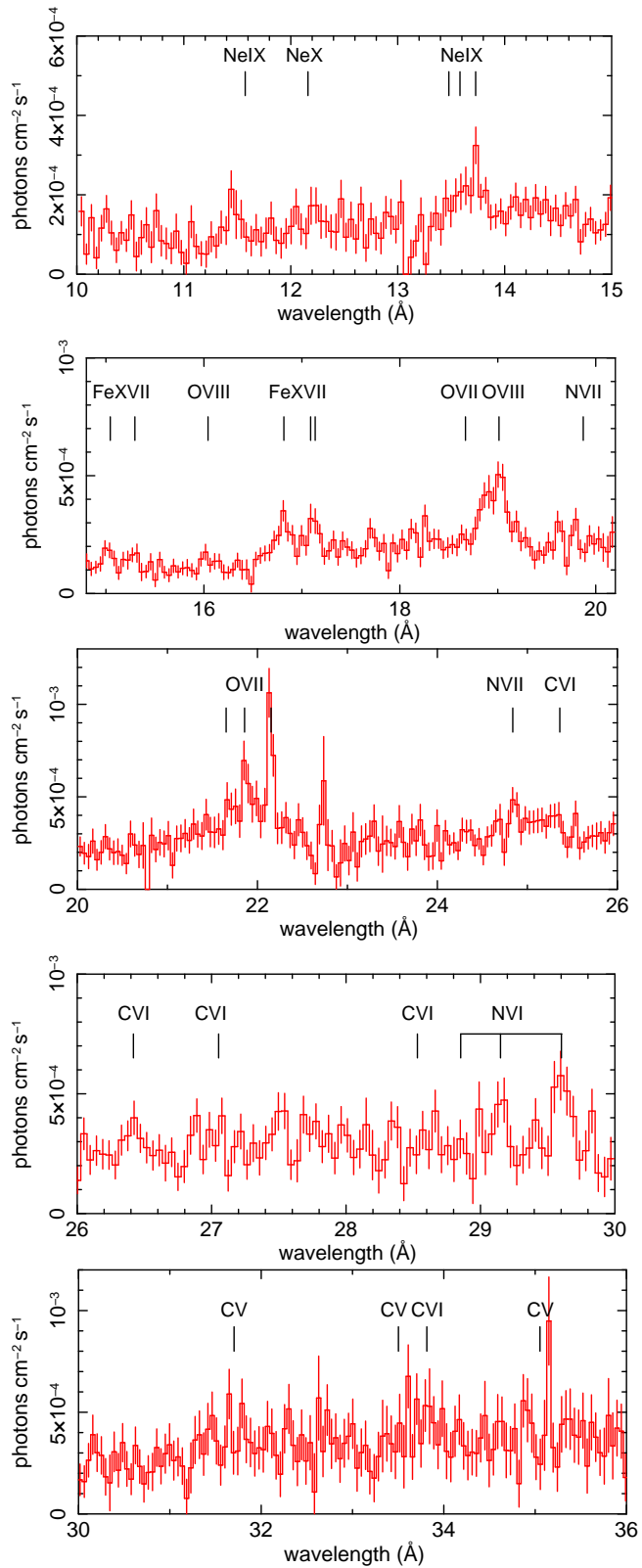
                                                          
\centering                                                              
\includegraphics[width=4.5cm, angle=270]{bl2.ps} 
\centering                                                              
\includegraphics[width= 4.3cm, angle=270]{bl1.ps}                     
\centering                                                              
\includegraphics[width=4cm, angle=270]{bl3.ps}                                                                                                 
\centering                                                              
\includegraphics[width=4.37cm, angle=270]{bl4.ps} 
\centering                                                              
\includegraphics[width=4.06cm, angle=270]{bl5.ps}                                                            
\caption                                                                
{Fluxed RGS spectrum of \ngc\ summed over 3 low-flux orbits, showing a dominant emission line spectrum at lower flux levels. The rest wavelengths  of
the main candidate emission lines are indicated, adjusted to the observing frame in the plot}      
\end{figure}

Figure 1 illustrates the combined and fluxed RGS spectrum summed over the three lowest-flux orbits (4,11,13 in figure 2) of the 2009  observation. The
emission spectrum is better defined than in any earlier observation of \ngc, benefiting from the long exposures and relatively low background prevailing
over the relevant part of each orbit. 

Among the more prominent features are narrow emission lines identified  with forbidden transitions in the He-like ions of NeIX (rest wavelength 13.698 \AA), OVII
(22.101 \AA) and NVI (29.534 \AA). In Paper I it was suggested that the narrow width and very low  velocity of the strong OVII   forbidden line might
indicate an origin in an accumulation of interstellar gas swept up by multiple  forward shocks.

In contrast, it was proposed there that the broad emission components, most obvious in figure 1 for OVIII Lyman-$\alpha$ (18.968 \AA)  and the region of
the OVII triplet ($\sim$21.6--22.1 \AA), but also visible in emission  lines of NeIX, FeXVII, NVII, NVI and CVI, might arise in a limb-brightened
shell  of  the cooling post-shock flow. We explore that concept further in the present paper.

\begin{figure}                                                                                                
\centering                                                              
\includegraphics[width=6.2cm, angle=270]{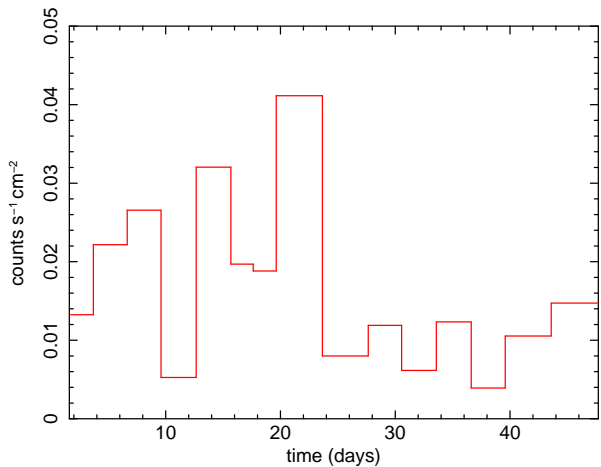}                                                                                  
\caption                                                                
{Mean RGS flux levels for the 15 \xmm\ orbits during the 2009 observation. The time axis is in MJD-54952}        
\end{figure}

Also evident in figure 1 is absorption at the core of several broad emission lines.  It is well known that absorption in a low velocity flow might be
underestimated due to `filling in' by re-emission  from an extended volume of the same ionised gas. Pounds \et\ (2004) suggested this was so for the
narrow, low velocity absorption lines dominating the  2001 \xmm\ observation of \ngc. The higher quality data from the 2009 observations now suggest the
interaction of emission and absorption is more complex.  

In the remainder of this paper we explore some general parameters of the broad line soft X-ray emission in \ngc\ by examining the velocity profiles of
OVIII Lyman-$\alpha$ and the OVII triplet (we outline the method of producing the velocity profiles in the Appendix). We then seek to constrain changes
in the ionisation state, and hence estimate the particle density in the post-shock flow, by a study of the OVIII Lyman-$\alpha$ velocity profile and of
several strong RRC,  as the mean continuum flux level changes.

\subsection{The velocity profile for OVIII Lyman-$\alpha$}

As noted above, the broad emission features are best seen in the low flux data, due to a higher contrast with the continuum, and minimal confusion of 
low velocity continuum absorption with the self-absorption in the broad lines.  We begin by examining the composite velocity
profile of the OVIII Lyman-$\alpha$ line for the 3 lowest flux observations (obs 4,11,13; hereafter the low3 data). 

Figure 3 (top) shows the resulting velocity profile, where the broad emission has been visually fitted with a positive Gaussian of width, velocity offset and amplitude
corresponding to that used in Paper I to better quantify absorption in the higher flux spectrum. (Although there is some indication that the actual
emission profile is more centrally peaked than a single Gaussian, as is also found in broad optical and UV lines, we retain the simpler model as a
satisfactory template throughout the present analysis). 
Represented in that way, the emission line is blue-shifted, with a centroid velocity v = -750 km s$^{-1}$ (where v=0 corresponds to the Lyman-$\alpha$  source
frame wavelength of 18.968 \AA), and 1$\sigma$  width of 1420 km s$^{-1}$. We note the measured width greatly exceeds the RGS resolution which corresponds, 
at $\sim$19 \AA,
to $\sigma$$\sim$400 km s$^{-1}$). 

While the single Gaussian is a good fit to the red wing and at the extreme part of the blue wing of the low3 data, strong absorption is evident at a
similar velocity to the peak of the emission line. It is important to note that the broad line emission is stronger than the continuum for these lowflux
data, with absorption being confined to low velocities. Allowance for the intrinsic RGS resolution confirms that the red wing emission is real, while
the absorption appears entirely to  the blue side  of zero velocity, a strong indication of self-absorption in the emitting gas. With that
interpretation the profile potentially  contains important information on the structure and flow geometry of the broad line gas, a point we take up in
Section 4. 

The mid panel of figure 3 adds the same template emission Gaussian to the composite profile for obs 9-15 (hereafter the low7 data), in the second  half of
the observation and all of relatively low flux. Again the fit to the red wing of the emission line is good, with the improved statistics sharpening the
onset of absorption on the blue side of zero
velocity. Continuum absorption is now seen near -5000 and -3500 km s$^{-1}$,
identified as separate high velocity outflow components in Paper I. 

The lower panel of figure 3 shows the composite OVIII Lyman-$\alpha$ velocity profile for all 15 data sets of the 2009 observation. The strong high
velocity outflow reported in Paper I is now seen in the absorption trough extending from $\sim$3000--7000 km s$^{-1}$, while a low velocity continuum absorption
component is apparent underlying the broad emission line. More directly relevant to the present study is the excellent visual fit of the lowflux emission
template to the all-data broad  emission line, again to the red wing and peak, and now also to the blue wing, supporting the working assumption in Paper I 
that the broad emission lines are quasi-constant throughout the 2009 observation.

In summary, we find the broad emission component of the OVIII Lyman-$\alpha$ line to be well described by a positive Gaussian of width 1$\sigma$ 
$\sim$1420 km s$^{-1}$ and centred at  -750 km s$^{-1}$, with strong absorption confined to the blue wing of the emission line. 

\begin{figure}
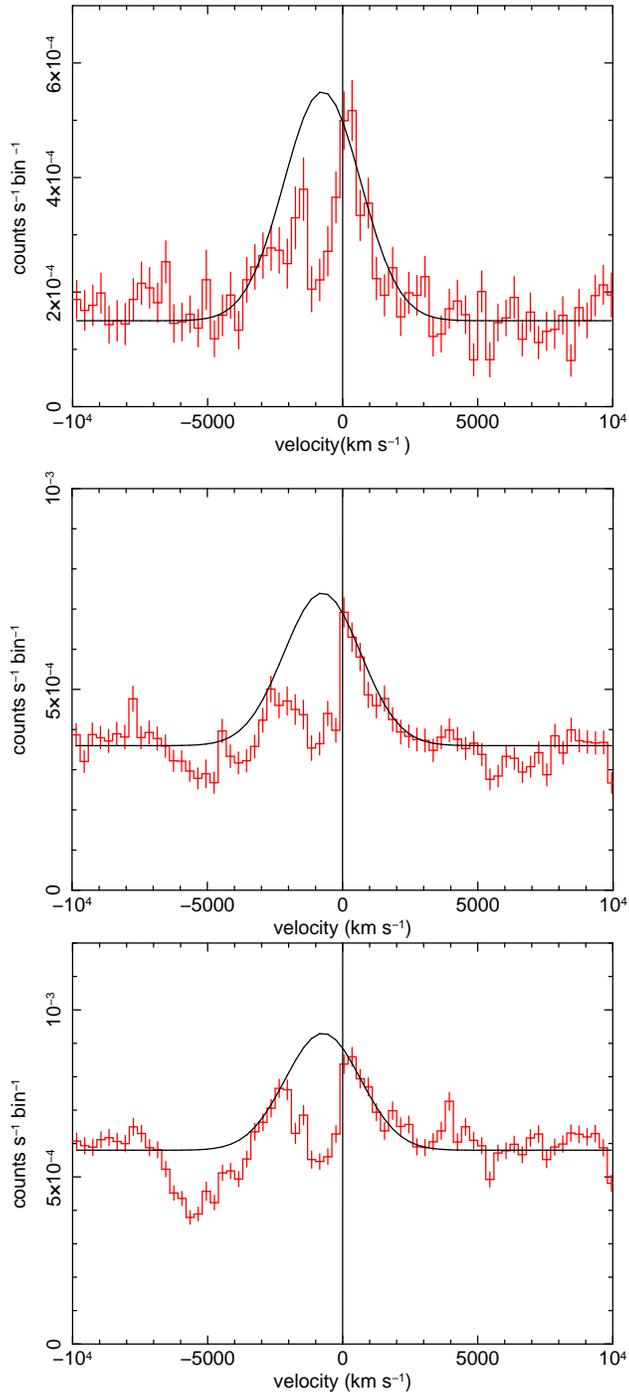
                                                                                                
\centering                                                             
\includegraphics[width=6cm, angle=270]{bl8.ps}                                                                                    
\centering                                                              
\includegraphics[width=6.4cm, angle=270]{bl12.ps}
\centering                                                             
\includegraphics[width=6cm, angle=270]{bl9.ps}                                                                                    
      
\caption                                                                
{Composite velocity profiles for the OVIII Lyman-$\alpha$ line. (top) Sum of the 3 lowest flux orbits, showing a broad and blue-shifted emission line
with  strong absorption near the line core. (middle) Sum of the 7 relatively low flux orbits in the second half of the observation. (lower) Velocity
profile for the complete 15 orbit observation. The positive Gaussian obtained from fitting the lowflux spectrum in Paper I (and  used in quantifying the
highflux absorption) is seen to match the broad emission line in all 3 composite velocity profiles, supporting the assumption in Paper I of a
quasi-constant emission arising from an extended region of ionised gas}        
\end{figure}

\subsection{The velocity profile for the triplet lines of OVII}

Broad line emission is also evident in the region of the OVII triplet (figure 1), and appears much broader than for OVIII Lyman-$\alpha$. A similarly
broad line was reported by Ogle \et\ (2004) and Steenbrugge \et\ (2009) and interpreted there as arising from an extreme high
velocity component of the BLR. We check later whether the improved quality of the new \xmm\ data allows resolution of the broad emission into 
individual triplet components, as might be expected, adopting individual component profiles as for OVIII Lyman-$\alpha$. However, as a
first step, we  model the overall emission of the OVII triplet with a single broad component, together with a narrow forbidden line.

Figure 4 (top panel) shows the OVII low3 velocity profile with the origin set at zero velocity for the intercombination line (rest wavelength
21.807\AA), close to the centre of the triplet. The forbidden line is found to be unresolved, with 1$\sigma$ width of 364$\pm$37 km s$^{-1}$ (RGS1 1$\sigma$
width $\sim$350 km s$^{-1}$ at 22\AA), and is blue-shifted by 180$\pm$37 km s$^{-1}$. The broad emission component is
strongly required, with $\chi$$^{2}$ falling from 146/86 dof (for the continuum plus 3 narrow triplet lines) to 120/87, with a 1$\sigma$ width of
3400$\pm$400 km s$^{-1}$, centroid velocity of -70$\pm$450 km s$^{-1}$ and amplitude 1.6$\pm$$0.2\times10^{-4}$ counts s$^{-1}$ bin$^{-1}$. We note that
excess counts in the red wing of the forbidden line suggest the fit might be affected by ignoring residual absorption in the OVI line (rest
wavelength 22.019\AA), found to be strong in Paper I.

The mid panel of figure 4 shows a further fit, with the addition of narrow resonance and intercombination line emission, and a  negative Gaussian to
allow for the OVI absorption. The outcome is a further improved fit ($\chi$$^{2}$/dof=96/81),  with the broad line width increasing to 4400$\pm$730 km
s$^{-1}$. The amplitude of the broad line is only marginally reduced by the inclusion of the narrow  emission  components, both of which appear to be
affected by absorption, the intercombination line probably by the OVI satellite line (21.79 \AA) and the resonance line by self-absorption. The result
is that both lines appear red shifted, in marked contrast to the forbidden line. The resonance line is most strongly  affected, with the observed
emission centroid at an unlikely  redshift of 470$\pm$140 km s$^{-1}$. We return to this in Section 2.3.

\begin{figure}
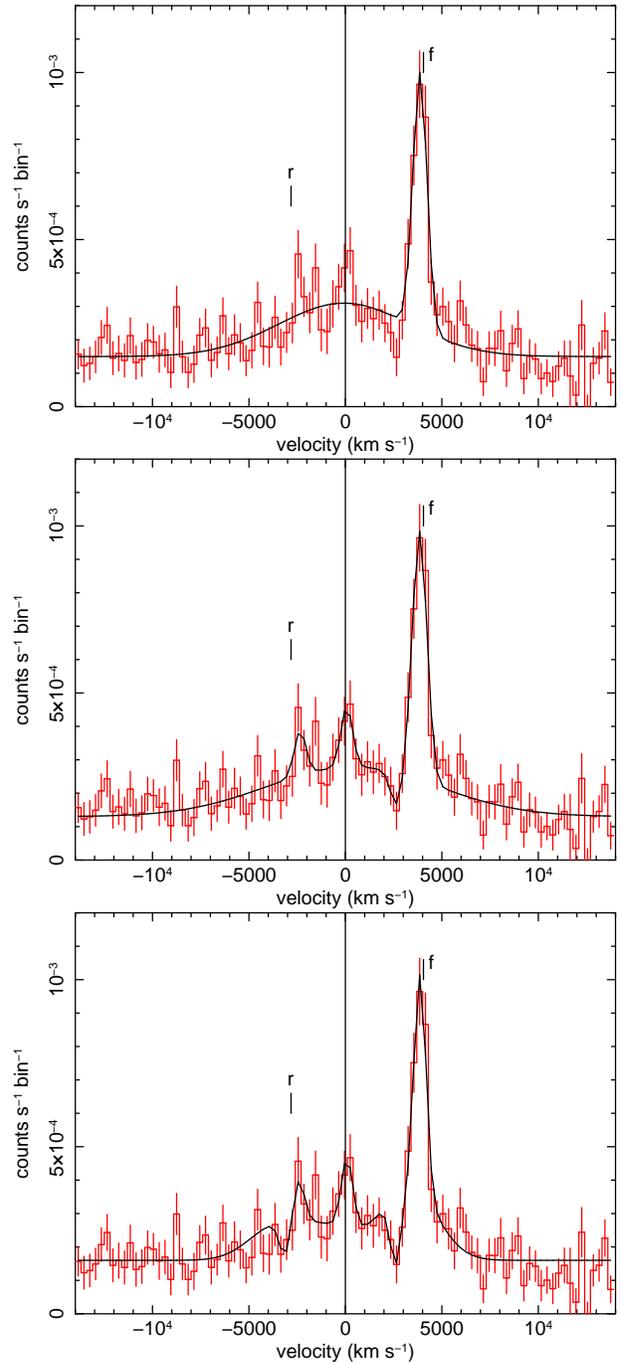
                                                                                             
\centering                                                              
\includegraphics[width=6cm, angle=270]{bl21a.ps}     
\centering                                                              
\includegraphics[width=6cm, angle=270]{bl21.ps} 
\centering                                                              
\includegraphics[width=6cm, angle=270]{bl21b.ps}                                                                                               
\caption                                                                
{Composite velocity profile for the 3 lowest flux orbits in the region of the OVII triplet, expressed in terms of a  velocity shift from the rest
wavelength of the intercombination line. The upper panel shows a fit with a single broad emission line in addition to the narrow forbidden line. The mid
panel is a statistically improved fit with the addition of  narrow intercombination and resonance emission lines and allowing for residual OVI
absorption near the blue wing of the forbidden line. The lower panel shows a further fit with separate broad emission components for the 3 triplet
lines}        
\end{figure}

On the basis of the above profile fitting, we find the OVII broad emission line is remarkably strong, the flux within the single broad line fit being
more than twice that in the forbidden line. However, the likelihood remains that the broad emission is a blend from all 3 components of the triplet,
assuming the origin is indeed velocity broadening. It would then be particularly interesting to resolve the separate components as their respective
strengths could be an important diagnostic of the density and mode of excitation of the emitting gas. 

To seek such a resolved profile we repeated the fit to the low flux data, replacing the single broad component with 3 separate components, each with a
blue shift and width  of -750  km s$^{-1}$ and 1420 km s$^{-1}$, the template from the fit to OVIII Lyman-$\alpha$. The amplitude of each line was left
free.  A further negative Gaussian was added to allow for absorption in the OVII resonance line. 

The main result of this further fit (figure 4, lower panel) was to quantify broad components of the separate triplet lines, finding   respective
amplitudes of 2.2$\pm$$0.5\times10^{-4}$ for the forbidden line, and 1.0$\pm$$0.3\times10^{-4}$ counts s$^{-1}$ bin$^{-1}$ for the resonance and
intercombination lines.  The narrow forbidden line velocity was unchanged and again unresolved, with a 1$\sigma$ width of 310$\pm$50 km s$^{-1}$. The
absorption lines of OVI and OVII yielded low outflow velocities consistent with the values obtained in Paper I.  

Although this more complex fit was statistically no better than the mid-panel fit, it did confirm the broad emission in OVII to be physically compatible
with that seen in OVIII Lyman-$\alpha$; moreover, finding the strongest broad component from the forbidden line  would rule out  a high density for the
emitting gas (and perhaps argue against the BLR origin proposed elsewhere).

\subsection{Implications of a strongly self-absorbed resonance line}

The diagnostic power of the He-like triplets in \ngc\ and other type 1 AGN has often been severely limited by strong absorption of the resonance line, a
situation we again find in the 2009 \xmm\ data. In particular, a dominant forbidden line cannot be taken as definitive evidence of a pure photoionised
gas (Porquet and Dubau 2000). The observation of a weak and  redshifted line in OVII suggests that the resonance line emission may be substantially
underestimated in \ngc, raising the possibility that the soft X-ray emission has a significant thermal component, which would be consistent with strong
two-body cooling in the later stages of the post-shock flow.  We also note that higher temperature (broader) RRC's would not be easy to resolve from the
continuum, even in spectra of the present quality.

While it is difficult to constrain the core OVII resonance emission in \ngc, the NVI triplet appears less strongly self-absorbed in the lowflux data
(Figure 5, top panel). The velocity profile for NVI, again centred on the intercombination line (29.082 \AA), shows an unresolved forbidden line (1$\sigma$ width
of 330$\pm$50 km s$^{-1}$) with a blue-shifted velocity of v = -140$\pm$50 km s$^{-1}$. The narrow intercombination line is weaker, but the red wing of the
resonance line is much better defined than for OVII.

In order to estimate the true strength of the resonance emission line, we repeated the NVI profile fit, including resonance and absorption lines
initially fixed  at velocities (relative to the source frame) of -180 km s$^{-1}$ and -400 km s$^{-1}$, respectively, taken from the measured positions
of the NVI forbidden and CVII Lyman-$\beta$  lines in the same plot. The outcome was a fit to the resulting P Cygni profile of the 
NVI
resonance line (figure 5, lower panel), with emission and absorption components of similar strength. In particular, the resonance emission amplitude, 2.4$\pm$$0.7\times10^{-4}$
counts s$^{-1}$ bin$^{-1}$, was now comparable to the forbidden line, and inconsistent with a pure photoionised gas. While photo-excitation will
increase the relative strength of the resonance line (and has been shown to provide a good fit to the soft X-ray spectra of NGC 1068, Kinkhabwala \et\
2002), a strong resonance line could indicate a significant thermal emission component in \ngc.

\begin{figure}
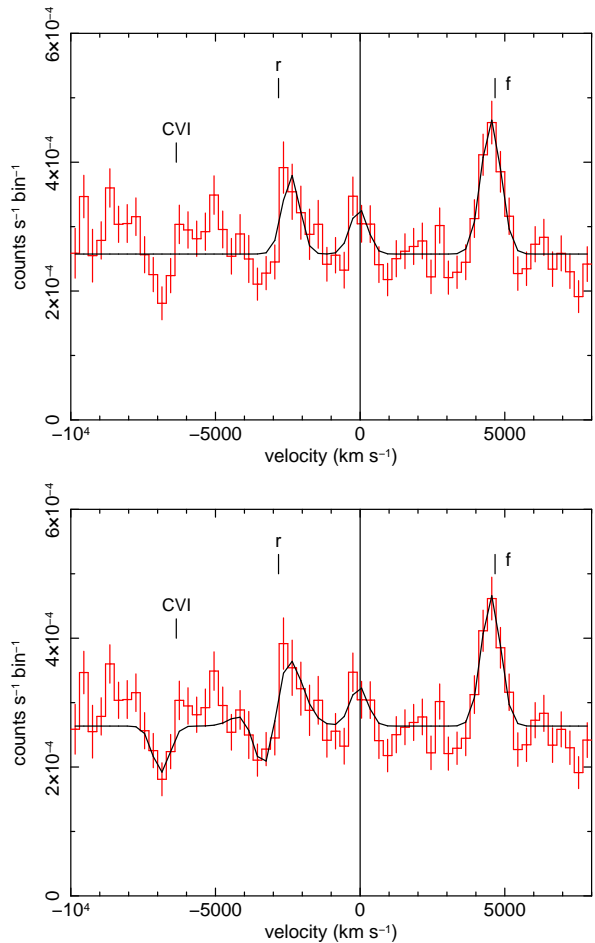
                                                                                                
\centering                                                              
\includegraphics[width=6.3cm, angle=270]{bl18a.ps}  
\centering                                                              
\includegraphics[width=6.3cm, angle=270]{bl18.ps}                                                                                      
\caption                                                                
{Velocity profiles for the NVI triplet from a composite of the low7 data, centred at zero velocity for the intercombination line. Zero velocity positions of
the forbidden and resonance lines are indicated as is that of CVI Lyman-$\beta$. The lower panel includes a PCygni fit to the NVI resonance line}        
\end{figure}

\section{Evidence for flux-linked variability in ionisation state of the shocked outflow}

The continuum absorption was found in Paper I to have a well determined velocity and ionisation structure. In terms of a cooling post-shock flow, the 
observed correlation of velocity and ionisation parameter requires the recombination time to be sufficiently short for the gas to adjust along the flow.
To check for that consistency requires the flow dimensions to be known, which in turn depends on breaking the degeneracy  between radial distance (from the
ionising source) and the particle density at some point along the flow.

Establishing a  flux-linked variability in the ionisation structure of the flow, and hence the related recombination rate and gas density,
is also of key importance in estimating the mass, energy and momentum of the outflow.  To pursue those aims we have examined the velocity profiles for each 
individual orbit of the 2009 observation. We focus here on profiles from observations of
relatively low mean flux  (and thereby less affected by absorption), and immediately following the highest-flux observation (obs 8 in figure 2). We again show the
profile centred on OVIII Lyman-$\alpha$, which also covers the region of the NVII RRC.

\begin{figure}
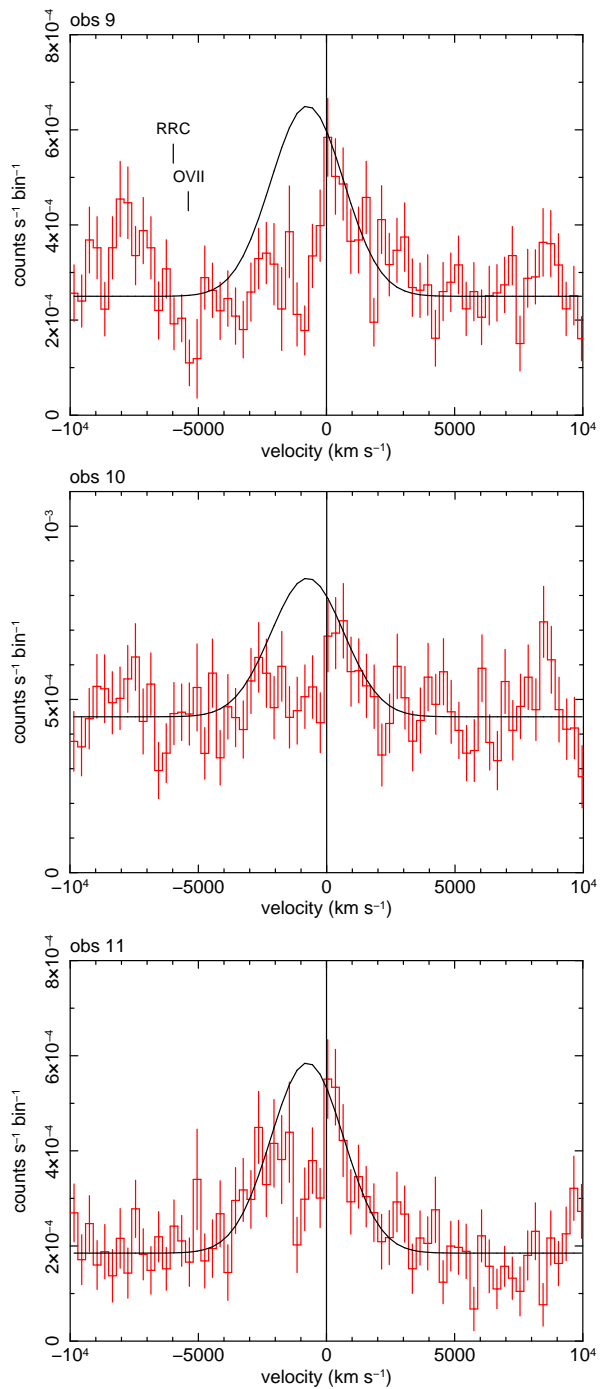
                                                                                                
\centering                                                              
\includegraphics[width=6.2cm, angle=270]{bl12a.ps}       
\centering                                                              
\includegraphics[width=6.05cm, angle=270]{bl12b.ps}    
\centering                                                              
\includegraphics[width=6.2cm, angle=270]{bl12c.ps}                                                                           
\caption                                                                
{Velocity profiles centred on the OVIII Lyman-$\alpha$ line for 3 successive orbits following several days of high continuum flux. A strong and
possibly blue shifted NVII RRC is seen to fade over several days}        
\end{figure} 

Figure 6 reproduces the profiles for obs 9,10 and 11. In each case the lowflux Gaussian template represents a satisfactory visual fit to the red 
(unabsorbed)  wing of the broad emission  line. To the blue side, only obs 11 (lower panel) is a good fit, with absorption more strongly affecting the
higher flux profiles. Nevertheless, to first order the broad line emission is essentially unchanged over the 3 successive orbits. 

However, the obs 9 profile  shows a strong excess to the blue side of v = -6000 km s$^{-1}$ (corresponding to the threshold wavelength of the NVII RRC).  
In obs 10, some 2 days later, the RRC is noticeably weaker, while in a further 4 days, by obs 11, it has faded from view. 

Although the RRC profile is not well determined, and might be affected by OVII absorption, an apparent blue shift of $\sim$2000 km s$^{-1}$ in the peak flux
is of particular 
interest, in suggesting an origin in the
intermediate  velocity outflow seen in absorption (Paper I). Importantly, if confirmed, that identification would link the intermediate velocity gas with a 
region in the flow where 
fully stripped nitrogen  is recombining over several days as the X-ray flux falls back from its peak in obs 8. A recombination timescale of $\sim$2-6
days for the relevant  component in the  outflow  would then correspond to a particle density  $\sim$$5\times10^{5}$cm$^{-3}$.

\subsection{Other RRC}

The potential importance of a variable NVII RRC, apparently responding to an earlier peak continuum flux, increases the interest in examining other RRC 
covered by the RGS data. The OVII RRC
(threshold wavelength 16.769 \AA) is found to be strong, and apparently variable, but its measurement is impeded by strong nearby FeXVII lines,  and by
a location at the long wavelength edge of the Fe UTA.  The OVIII and NeIX RRC fall in regions where the RGS sensitivity is falling. However, both CV and CVI 
are found to exhibit strong RRC, and are less
affected by other spectral features. 

The top two panels of figure 7 show velocity profiles for the CVI RRC, centred at the threshold wavelength of 25.303 \AA\ (note the binning  here is 600
km s$^{-1}$, with an extended range to the blue side to cover the region of the NVII Lyman-$\alpha$ line at 24.781 \AA\  and to better determine the 
continuum level). An emission line template is added visually to the top panel and carried over to the higher flux level profile in
the middle panel, the primary aim here being to show that the CVI RRC is well resolved. 

As with the NVII RRC, the peak CVI RRC flux is seen to
shift to the blue as the continuum level increases, and in this case there are no likely absorption lines to affect the RRC profile close to the threshold
wavelength.
The similarity in the CVI and NVII RRC blue shifts at higher
continuum  levels strengthens a direct association of the recombining plasma with the strong intermediate velocity absorption ($\sim$3500-5000 km s$^{-1}$) 
observed 
in the
same higher level ions (Paper I). 

Furthermore, tracking the CVI RRC through individual orbits shows a similar pattern to the NVII RRC, again indicating a response time for the intermediate 
velocity
gas of a few days. In contrast, the lower level CV RRC (31.63 \AA) shown in figure 7 (lower panel) only appears at the zero velocity threshold, 
consistent with the low absorption velocities characteristic for
that ion. 

In summary, we find RRC of NVII and CVI to vary in strength and velocity profile over several days and in a manner apparently dependent on the continuum flux
level. We interpret the variability as from enhanced photoionisation of the high velocity flow when the continuum flux level is high, being followed
by  strong recombination over the following days of reduced flux level. The velocity and ionisation  gradients found in the absorption spectra (Paper I) then 
explain the differences between the RRC of NVII, CVI and the lower ionisation state of CV.  

The best determined RRC profiles all indicate a relatively low temperature of $\sim$7$\pm$2 eV.

\begin{figure}
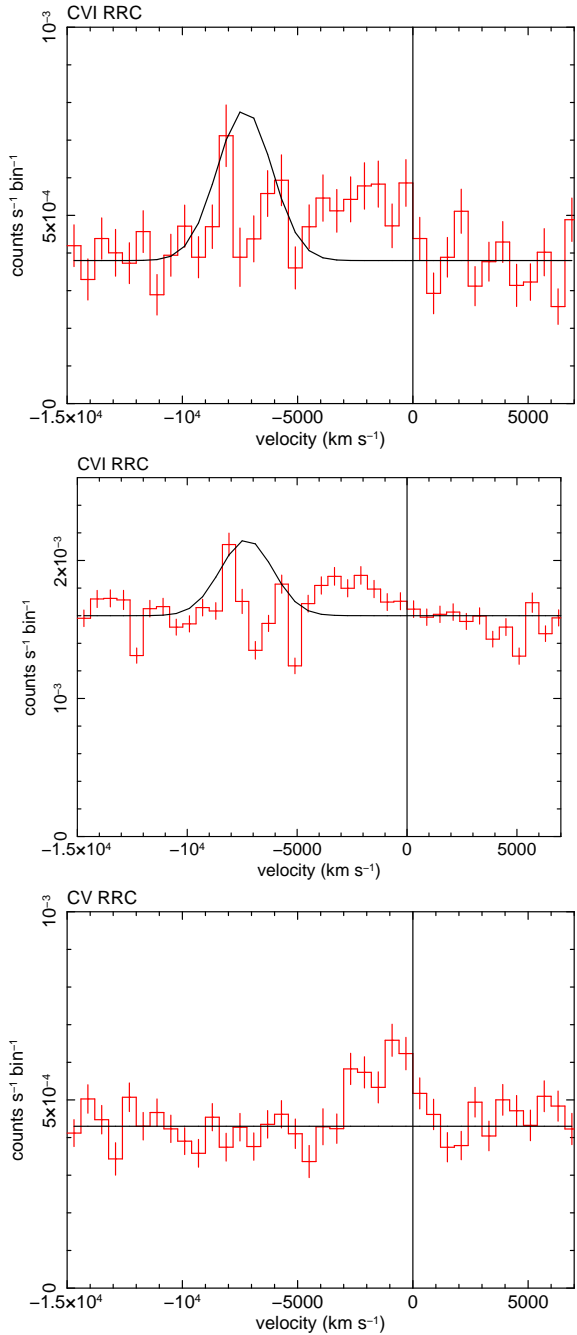
                                                                                                
\centering                                                              
\includegraphics[width=6cm, angle=270]{CVI_RRC.ps}
\centering                                                              
\includegraphics[width=5.7cm, angle=270]{CVI_RRC_a.ps}                                                                                     
\centering                                                              
\includegraphics[width=6cm, angle=270]{CV_RRC.ps}                                                                                     
                                                                                     
\caption                                                                
{(top and mid panels) Velocity profiles of the CVI RRC for two different continuum levels. Also shown is the template Gaussian fit to the NVII Lyman-$\alpha$
line clarifying the resolution of the RRC. The peak RRC flux is seen to shift to the blue
for higher continuum levels, as also found for the NVII RRC. (lower panel) The CV RRC is always seen at the zero velocity threshold}        
\end{figure}

\section{Discussion}

In Paper I we reported a rich absorption line spectrum from the 2009 \xmm\ observation of \ngc, revealing the presence of an
ionised outflow with a wide range of velocities and ionisation parameter.  The absorption line velocity structure and a broad  correlation of
velocity with ionisation parameter were shown there to be consistent with an outflow scenario where a  highly ionised, high velocity wind runs into the
interstellar medium  or previous ejecta, losing much of its kinetic energy in the resultant strong shock (King 2010). With the strong immediate post-shock cooling likely
to be dominated
by Compton scattering of the  AGN thermal continuum (King 2003), we also noted that a quasi-constant soft X-ray emission component might be evidence
of further
energy loss as the post-shock gas slowed and recombined ahead of the contact discontinuity.

A second outstanding feature of the soft X-ray data from the 2009 observation was a complex emission line spectrum, particularly evident at low continuum flux levels, with velocity-broadened emission from 
several H-
and He-like resonance lines, as well as a number of strong RRC. 
Broad emission lines of OVII and OVIII have been reported previously from \ngc\ (Ogle 2004, Steenbrugge 2009), and attributed to scattering of the AGN X-ray continuum 
from high velocity clouds in the BLR. An alternative  interpretation, outlined here,
envisages the broad emission lines arising from the limb-brightened shell of shocked gas building up ahead of the contact discontinuity. On this alternative picture the line  
broadening primarily arises from the angular divergence of the flow at the bright limb of the expanding spherical shell. 

An optically thin spherical shell would produce an emission line centered on zero velocity (in the AGN rest frame). Since in a spherical geometry the 
absorbing gas will always have a positive line-of-sight velocity relative to the emission, self absorption in the shell will set the onset of absorption
at zero velocity.  The latter is precisely what the data (eg figure 3) show, although the peak of emission is shifted to the blue.

Figure 8 sketches the geometry of such a limb-brightened shell, where the broad emission line of OVIII Lyman-$\alpha$ is depicted arising from a
near-orthogonal component of
the high velocity radial flow, seen in absorption in line of sight to the continuum, but now with lower apparent velocities observed in emission from the 
limb.
In the figure, for a typical post-shock absorption velocity of 5000 km s $^{-1}$ (Paper I) the red and blue wings (at HWHM) of the
OVIII Lyman-$\alpha$ line correspond to rear and forward inclinations a$\sim$12$\deg$ and b$\sim$30$\deg$, respectively.

While the blue offset in the observed broad emission line profiles might be an artefact of the self
absorption, it is perhaps more likely evidence of a dominantly near-sided/approaching outflow.
However, we also note the red wing in the OVIII Lyman-$\alpha$ line is broader than the instrument
response, indicating that at least part of the far, receding side of the shell is visible.  

The putative torus is an obvious candidate for obscuring part
of the receding flow in the shell, given the estimated shell radius of $\sim$$7\times10^{17}$ cm (Paper I). 
In that context it is interesting to note that
a strong blue-shifted asymmetry in the [OIII] line profile of \ngc\ was described by Veilleux (1991) to `favour radial motion and a source of
obscuration'. 
On still larger scales, a narrower, forward projected conical
outflow was observed in [OIII] imaging and spectroscopy of \ngc\ (Christopoulou \et\ 1997).

\begin{figure}                                                                                                
\centering                                                              
\includegraphics[width=6.5cm, angle=0]{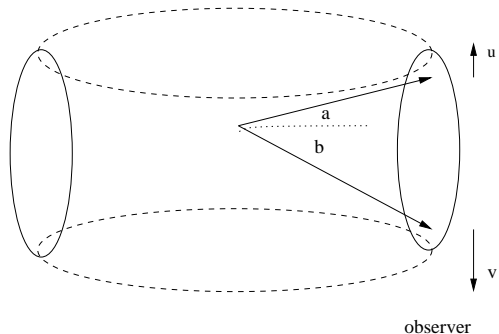}                                                                                  
\caption{Geometry of a limb-brightened shell, where the broad line of OVIII Lyman-$\alpha$ is explained as emission from the high
velocity radial flow, seen in continuum absorption when in line of sight to the nucleus. The indicated flow lines correspond to the red (u) 
and blue (v) wings of the emission line
profile when viewed by the observer. The central offset corresponds to a dominantly near-sided/approaching outflow}        
\end{figure}

With the flow of gas in the limb brightened shell near-orthogonal to the line of sight, the geometry of figure 8  also offers a natural explanation 
of the separate low velocity absorption component, found in Paper I to occur over a wide range of ionisation, in contrast to the higher velocity 
absorption where the
velocity and ionisation parameter are strongly correlated.  

In contrast to the quasi-constant broad emission lines, the RRC of NVII and CVI are found to vary, in both strength and profile, on a timescale of days. 
Given that the change appears to
respond to the continuum flux level, we interpret the variability in terms of the recombination time and hence derive a particle density 
for the relevant intermediate
velocity/ionisation region in the post-shock flow of  order $5\times 10^{5}$ cm$^{-3}$.  

The relatively short recombination time provides an important consistency check on the observed velocity/ionisation 
gradient in the absorption spectra reported in Paper I, and shown there to be a simple consequence of mass conservation. In particular, that outcome
requires the flow to recombine sufficiently quickly as it decelerates, in order to maintain the observed velocity/ionisation correlation. 
With an estimated shell thickness (Paper I) of
$\sim$$4\times 10^{15}$ cm, and for a mean velocity of $\sim$4000 km s$^{-1}$, the shell transit time of a few months is comfortably longer than the recombination
time along the flow.    
We also note that the relatively low density in the post-shock gas is consistent with finding a broad component to the OVII forbidden line.  

Finally, while the strong RRC are indicative of a photoionised gas, as is the low temperature kT$\sim$7eV, we also find a narrow resonance emission line in the
NVI triplet comparable in strength to the narrow forbidden line, which could imply an electron temperature an order of magnitude higher (fig. 7 in Porquet and Dubau,
2000) and indicate significant thermal emission. 

However, some of the clearest soft X-ray emission spectra to date, from observations of the bright Seyfert 2 galaxies, Mkn 3 (Sako \et\ 2000) and NGC 1068 
(Kinkhabwala \et\ 2002, Ogle \et\ 2003), have shown that the X-ray
emission in those objects is explained by radiative recombination and decay in an outflow which is photoionised and  photoexcited by the central
continuum, with no significant thermal emission.

Nevertheless, the obscuring torus in such a type 2 object is likely to hide emission from the innermost, high velocity flow.  
In that respect it is interesting to recall that \xmm\ EPIC spectra  of
a sample of 2MASS sources suggested the `hidden' soft X-ray emission could be relatively strong (Pounds and Wilkes 2007, Wilkes \et\ 2008), while we also note 
that the high quality soft X-ray spectra of type 2 Seyferts show no evidence for broad line emission. 

In that wider context, the 2009 \xmm\ observation
of \ngc\ may offer one of the best opportunities to date to study the full soft X-ray emission spectrum of a Seyfert galaxy. On our interpretation, at
least, the inner regions are of special interest in harbouring shocks that are likely to strongly affect the dynamics of the subsequent
outflow and its interaction with the host galaxy.

\section{Summary}

Broad emission lines of OVIII Lyman-$\alpha$ and the OVII triplet, possibly including a broad component of the forbidden line, are found to be strong
throughout  the 2009 \xmm\ observation of \ngc. 

The low centroid velocity of the OVIII Lyman-$\alpha$ broad line is consistent with an origin in the limb-brightened shell of post-shock gas, with emission
from the near-orthogonal component of the flow seen  in continuum
absorption at $\sim$5000 km s$^{-1}$.

The same limb-brightened geometry provides a natural explanation for the sharp onset of absorption to the blue side of zero velocity, found in all individual velocity
profiles, being interpreted as a consequence of the radial flow through the limb-brightened shell leading all absorbing gas to have a  blue shift
relative to the source of emission.

The detection of RRC of NVII and CVI which vary in both flux and velocity profile, as the continuum level changes, has yielded an important measure of the
recombination timescale for ionised gas in the related high velocity flow. Breaking the degeneracy between radius and particle density, that so often limits the assessment
of AGN outflows, has allowed the radius
and thickness of the shell of post-shock gas to be estimated, and in turn has confirmed that the high velocity flow is able to recombine sufficiently fast to
maintain the observed correlation of ionisation parameter and velocity reported in Paper I.

Strong low velocity absorption renders the OVII resonance line essentially unconstrained, and leads to narrow components of OVIII Lyman-$\alpha$ and
other resonance lines in \ngc\ (and perhaps more generally in type I Seyferts) to be poorly determined. Together with the observational difficulty of detecting much 
broader (and hotter) RRC, we conclude that a significant thermal
contribution to the soft X-ray emission of \ngc, cannot be ruled out. 
Nevertheless, with the relatively low particle densities in the post-shock flow, two-body cooling is likely to remain less important than Compton cooling which
we have argued (Paper I, King 2003) will be strong so close to the AGN continuum.

\section*{ Acknowledgements} 
The results reported here are based on observations obtained with \xmm, an ESA science mission with instruments and contributions directly
funded by ESA Member States and the USA (NASA). The authors wish to thank Andrew King and Mike Goad for useful discussions, the referee for a careful reading
of the manuscript, and  the SOC and SSC 
teams for organising  the \xmm\ observations  and initial data reduction.

\section{Appendix}

In this section we describe the method used to produce the line velocity profiles which are the basis of the analysis in this paper.
We choose the velocity plots as they provide a convenient (and accurate) way of combining different data sets, and present the parameters
of interest in the directly relevant velocity coordinates.
Each RGS observation was first
processed using {\tt RGSPROC} to produce a set of first-order source and background spectra, along with appropriate response
matrices, from each of RGS1 and RGS2, with $3400$ wavelength channels.

The simplest way to produce a velocity profile for several lines from several observations is to sum the counts in each spectral
channel across the observations, transform the wavelengths of the channels around each line of interest to velocity shifts relative
to the (source-frame) wavelength of the line, and then average (or sum) over the different lines. Unfortunately this simple
procedure is complicated by three facts about the RGS data. Firstly, the RGS effective area can change sharply due to bad pixels,
columns or even whole CCDs. Secondly, small differences in pointing between the observations mean that the wavelength-channel
conversion differs slightly between observations. (At least, this is true for the standard products produced by {\tt RGSPROC}). And
thirdly, the wavelength resolution of the RGS is approximately constant across the spectral range, meaning that the velocity
resolution differs for lines at different wavelengths. Taken together the first two facts mean that co-adding spectra by summing
the counts in each channel will result in a degradation (blurring) of the line response function, and may produce spurious
line-like features (due to narrow ``drop outs'' in the spectra) if not corrected using an effective area curve computed for the
merged data. The third point means that the velocity bins for one line will not be well matched to those of other lines, making it
difficult to average the counts in velocity bins over different lines.

The solution we employ is to convert each spectrum from counts per channel, with a corresponding channel to wavelength range
conversion, to a list of wavelengths for each event. We do this by randomising the wavelength of each event in a given spectral
channel, in a given observation, within the wavelength range of the channel specified in the appropriate (observation-specific)
response matrix. This effectively `unbins' the spectrum, but as the wavelength range of each channel (in the $3400$ channel RGS
spectral products) is far smaller than the resolving power of the RGS, the randomisation within a given channel leads to negligible
degradation of the spectral resolution. The wavelengths of the counts around specific lines of interest can be converted to
velocity shifts (after corrected for the source's systemic redshift), and the counts rebinned into velocity bins of width $\Delta
v$. The counts per velocity bin are then summed over different observations and different lines to produce a composite line profile
in velocity space. Error bars are computed assuming simple counting statistics (i.e Poissonian fluctuations within each velocity
bin). The same procedure is applied to the background spectra to produce a composite background spectrum in velocity space. 

The complex changes in effective area as a function of wavelength mean that the relative weighting of source and background spectra
is itself a complex function of wavelength. The effective exposure time per velocity bin of the composite was calculated by summing
the product of the individual exposure times of the observations and a wavelength-dependent quality function to account for `drop
outs' in individual spectra (caused by bad CCD columns or missing chips, which contributed zero effective exposure for certain
wavelengths). This was converted to velocity space (around the lines of interest), averaged in velocity bins, summed over the
different lines. The result is the average exposure time each velocity bin acquired from each observation and each line
contributing to the spectrum. This was then used to calculate the effective count rate (ct s$^{-1}$) per velocity bin for both
source and background spectra.

The wavelength-dependent scaling of source and background extraction regions was calculated by  summing the effective area curves
(as a function of wavelength) of each observation, weighted by the observational exposure time, separately for the source and
background data. The ratio of the resulting curves gives the source/background scaling as a function of wavelength. This scaling
function was then converted to velocity shifts (around the lines of interest), averaged in velocity bins, summed over the different
lines, and the result used to perform the velocity-dependent background subtraction. The final result is a background-corrected
count rate as a function of velocity.

\end{document}